\documentclass[Proceedings]{SciPost2}
\usepackage{graphicx}
\usepackage{subcaption}
\usepackage{tabu}

\usepackage{url}
\usepackage{hyperref}
\definecolor{nicered}{rgb}{0.5,0.,0.}
\definecolor{nicegreen}{rgb}{0.,0.5,0.}
\definecolor{niceblue}{rgb}{0.,0.,0.5}
\hypersetup{colorlinks,citecolor=nicegreen,linkcolor=nicered,urlcolor=niceblue}

\usepackage{amsmath,amssymb}
\usepackage{physics}
\DeclareSymbolFont{usualmathcal}{OMS}{cmsy}{m}{n}
\DeclareSymbolFontAlphabet{\mathcal}{usualmathcal}

\DeclareSymbolFont{usualmathcal}{OMS}{cmsy}{m}{n}
\DeclareSymbolFontAlphabet{\mathcal}{usualmathcal}

\begin{document}
	
	{\hfill MSUHEP-21-017}

\begin{center}{\Large \textbf{
Connected and Disconnected Sea Partons from CT18 Parametrization of PDFs\\
}}\end{center}

\begin{center}
Tie-Jiun Hou\textsuperscript{1,*},
Jian Liang\textsuperscript{2},
Keh-Fei Liu\textsuperscript{3},
Mengshi Yan\textsuperscript{4},
and
C.--P. Yuan\textsuperscript{5}
\end{center}

\begin{center}
{\bf 1} Department of Physics, Northeastern University, Shenyang 110819, China
\\
{\bf 2} Guangdong Provincial Key Laboratory of Nuclear Science, Institute of Quantum Matter, South China Normal University, Guangzhou 510006, China \\
Guangdong-Hong Kong Joint Laboratory of Quantum Matter, Southern Nuclear Science Computing Center, South China Normal University, Guangzhou 510006, China 
\\
{\bf 3} Department of Physics and Astronomy, University of Kentucky, Lexington, KY 40506, U.S.A.
\\
{\bf 4} Department of Physics and State Key Laboratory of Nuclear Physics and Technology, Peking University, Beijing 100871, China
\\
{\bf 5} Department of Physics and Astronomy,
Michigan State University, East Lansing, MI 48824, U.S.A.
\\
* tjhou@msu.edu
\end{center}

\begin{center}
\today
\end{center}

\definecolor{palegray}{gray}{0.95}
\begin{center}
\colorbox{palegray}{
  \begin{tabular}{rr}
  \begin{minipage}{0.1\textwidth}
    \includegraphics[width=22mm]{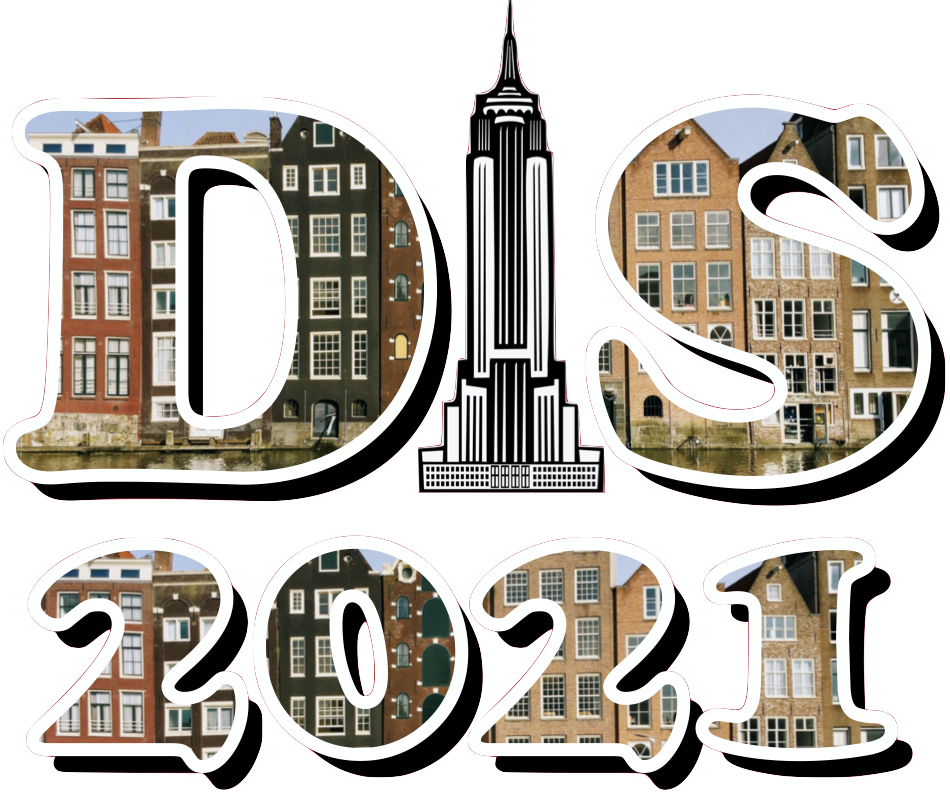}
  \end{minipage}
  &
  \begin{minipage}{0.75\textwidth}
    \begin{center}
    {\it Proceedings for the XXVIII International Workshop\\ on Deep-Inelastic Scattering and
Related Subjects,}\\
    {\it Stony Brook University, New York, USA, 12-16 April 2021} \\
    \doi{10.21468/SciPostPhysProc.?}\\
    \end{center}
  \end{minipage}
\end{tabular}
}
\end{center}

\section*{Abstract}
{\bf
The separation of the connected and disconnected sea partons, which were uncovered in the Euclidean path-integral formulation of the hadronic tensor, is accommodated with the CT18 parametrization of
the global analysis of the parton distribution functions (PDFs). This is achieved with the help of the distinct small $x$ behaviors of these two sea parton components and the constraint from the lattice calculation of the ratio of the strange momentum fraction to that of the ${\bar u}$ or ${\bar d}$ in the disconnected insertion. This allows lattice calculations of separate flavors in both the connected and disconnected insertions to be directly compared with the global analysis results term by term.
}


\section{Introduction}
\label{sec:intro}

Extracting PDFs information is intrinsically an inverse problem, since the factorization formula involves an integral of the product of the parton distribution functions (PDFs) and the perturbative short distance kernel. 
The common approach is to
model the PDFs in terms of the valence and sea partons with respective small and large $x$ behaviors
and perform a global fit of the available experimental data at different $Q^2$ with evolution. 
In paricular, the flavor structure of the partons and antipartons can be improved with experiments which directly address the flavor dependence. For example, the NMC measurement~\cite{nmc} of $\int^1_0 dx [F^p_2(x)-F^n_2(x)]/x$ turns out to be $0.235 \pm 0.026$, a 4 $\sigma$ difference from the Gottfried sum rule $I_G \equiv \int^1_0 dx [F^p_2(x)-F^n_2(x)]/x  =1/3$, which implies that the $\bar u = \bar d$ assumption was invalid~\cite{Gottfried:1967kk}.

The violation of the Gottfried sum rule prompted the Euclidean path-integral formulation of the 
the hadronic tensor of the nucleon which uncovered that there are two kinds of sea partons, one is the
connected sea and the other disconnected sea~\cite{Liu:1993cv,Liu:1999ak}. The connected sea (CS) results from a connected insertion of the currents on the `valence' quark lines and the disconnected sea (DS) is from a disconnected insertion involving a vacuum polarization from the quark loop involving  the external currents. It was proven~\cite{Liu:1993cv} that, in the isospin symmetric limit, the Gottfried sum rule violation originates only from the CS which is subject to Pauli blocking due to the unequal numbers of the valence $u$ and $d$ quarks in both the proton and the neutron. 

\begin{figure}
    \centering
    \includegraphics[width=0.32\textwidth]{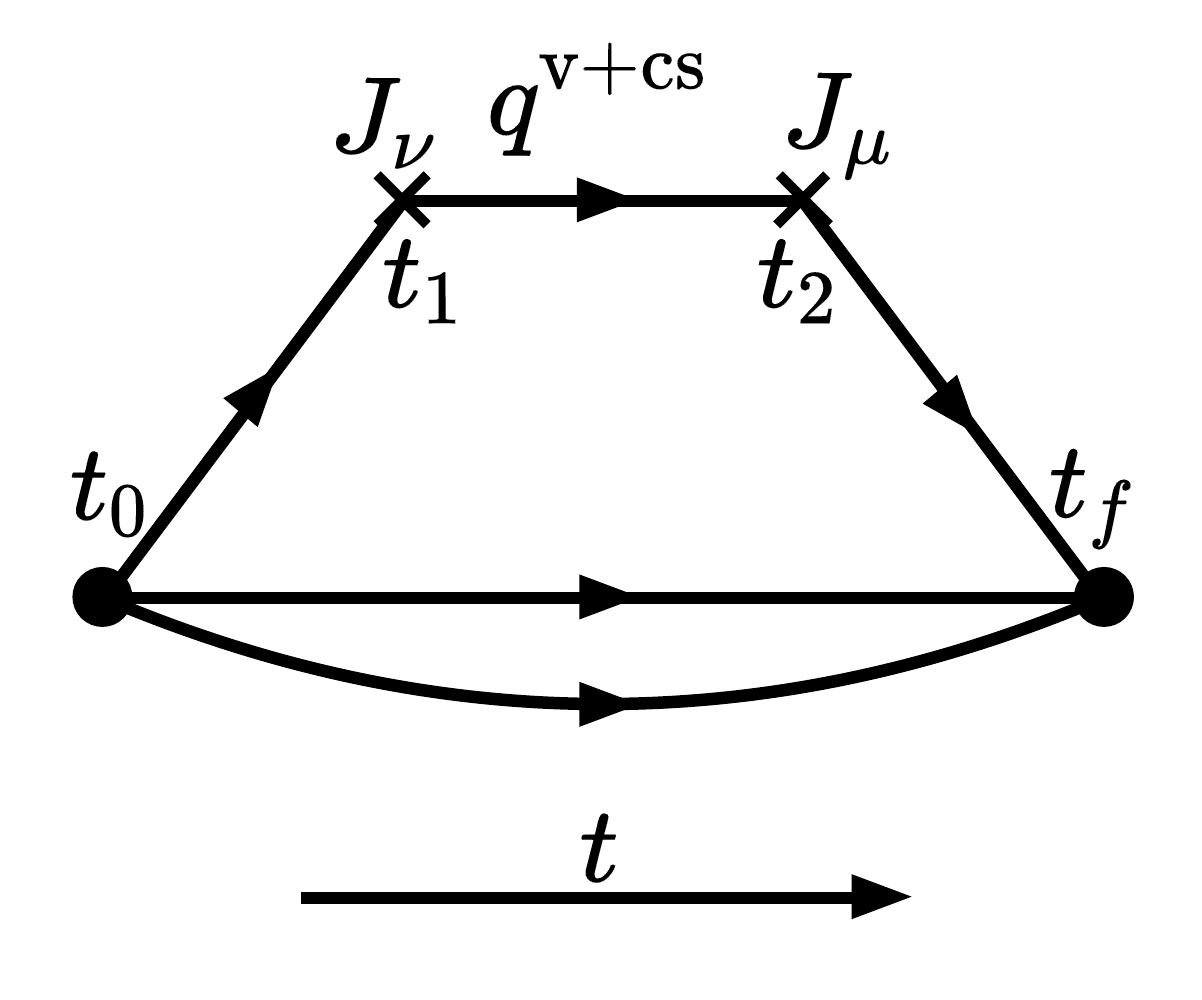}
    \includegraphics[width=0.32\textwidth]{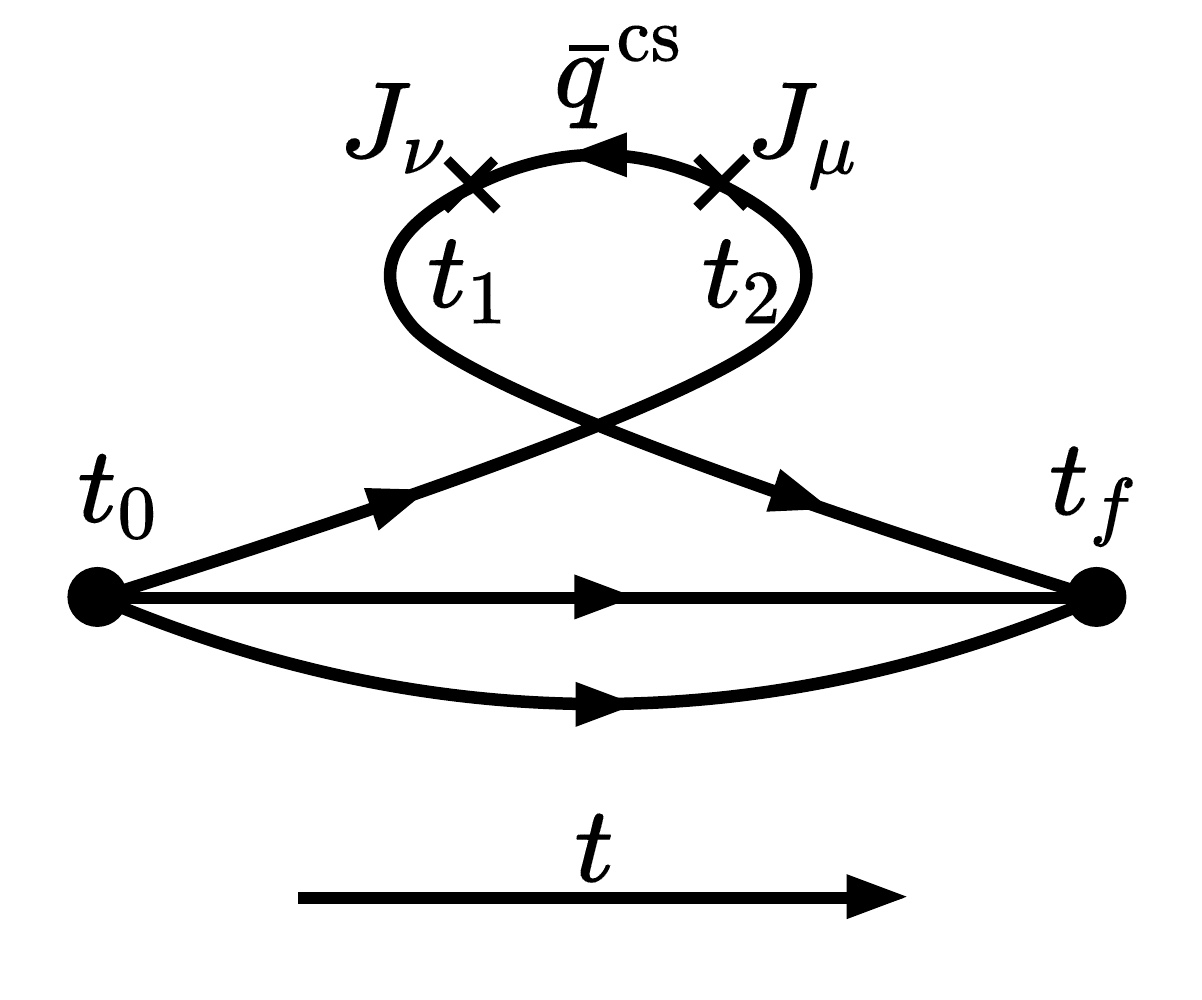}
    \includegraphics[width=0.32\textwidth]{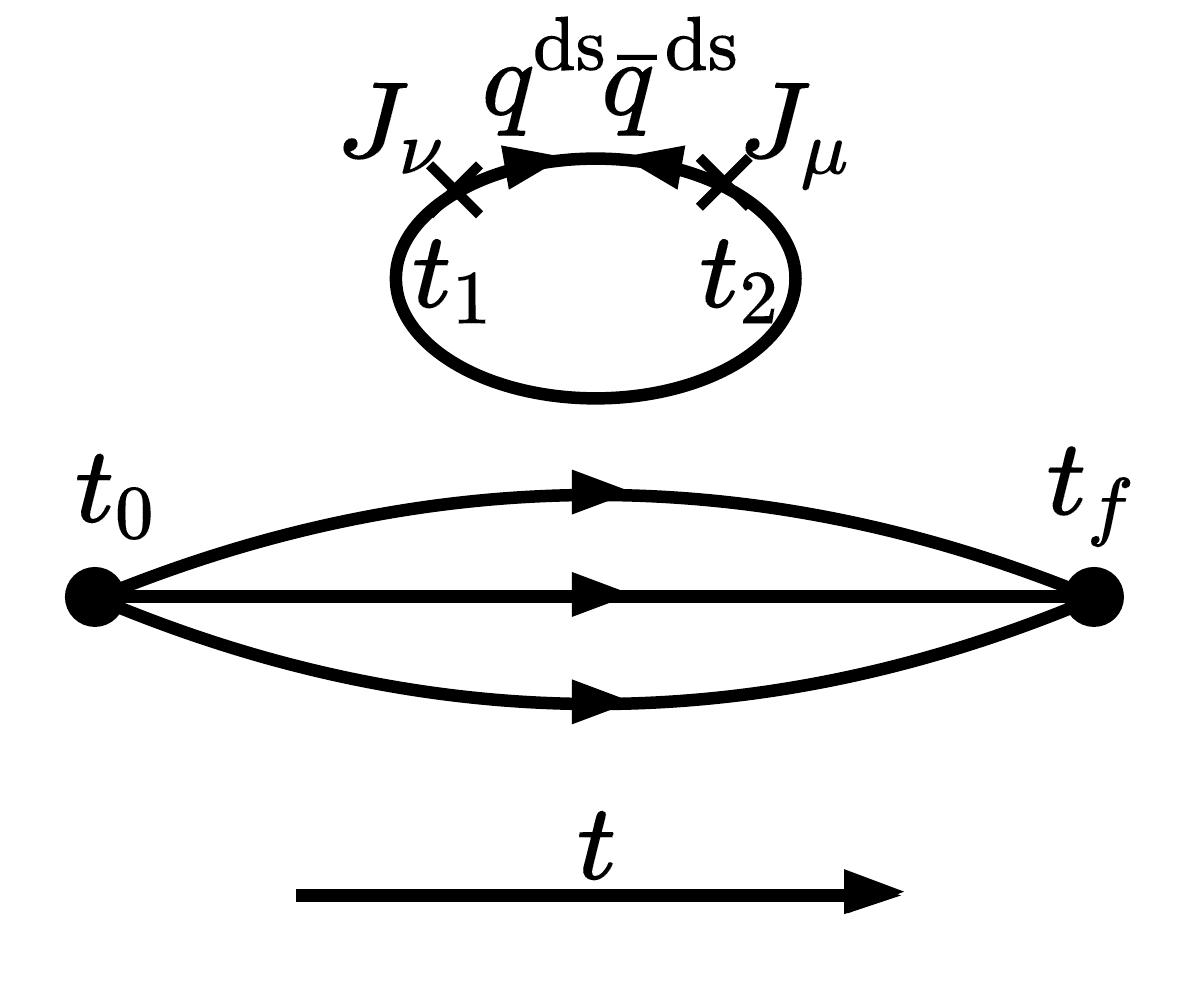}
    \caption{Three gauge invariant and topologically distinct insertions in the Euclidean-path integral
formulation of the nucleon hadronic tensor where the currents couple to the same quark
propagator. In the DIS region, the parton degrees of freedom are as follow.
  The left panel: the valence and connected sea (CS) partons $q^{v+cs}$; the middle panel: the CS anti-partons $\bar{q}^{cs}$; the right panel: the disconnected sea (DS) partons $q^{ds}$ and anti-partons $\bar{q}^{ds}$ with $q = u, d, s,$ and $c$. Only $u$ and $d$ are present in the left and middle panels for the nucleon hadronic tensor.\label{leading-twist}}
\end{figure} 

In this work, we shall accommodate parton degrees of freedom delineated in the path-integral formulation of the hadronic tensor in the form of CT18 global analysis \cite{Hou:2019efy} of unpolarized PDFs.

 \section{Global fitting} \label{sec:GlbFit}

As a common practice, the global fitting programs adopt the parton degrees of freedoms as $u, d, \bar{u}, \bar{d}, s, \bar{s}$ and $g$ at the initial scale $Q_0$, about 1 GeV, from where the PDFs are evolved via DGLAP equations\cite{Altarelli:1977zs,Dokshitzer:1977sg,Gribov:1972ri}. However, we see that from the path-integral formalism of QCD, each of the
$u$ and $d$ have two sources, one from the connected insertion (CI) (left panel of Fig.~\ref{leading-twist}) and one
from the disconnected insertion (DI) (right panel of Fig.~\ref{leading-twist}), so are $\bar{u}$ and $\bar{d}$ from the middle and right panels of Fig.~\ref{leading-twist}. On the other hand, $s$ and $\bar{s}$ only come from the DI (right panel of Fig.~\ref{leading-twist}).
In other words,
\begin{align}  \label{dof}
      &u=u^{v+cs} + u^{ds},             &   & d= d^{v+cs} + d^{ds} \nonumber \\
&\bar{u}=\bar{u}^{cs} + \bar{u}^{ds},   &   & \bar{d}= \bar{d}^{cs} + \bar{d}^{ds}, \nonumber \\
      &s=s^{ds},                        &   & \bar{s}= \bar{s}^{ds},
\end{align}
In CT18 \cite{Hou:2019efy}, at $Q_0=1.3$ GeV, the non-perturbative PDFs parametrization involves totally 6 degrees of freedom: $g, u^v, d^v, \bar{u}, \bar{d}$, and $s$, with the  assumption that $s = \bar{s}$.
When the separation of CS and DS partons are considered, we would have more partonic degrees of freedom (11 in total) at the $Q_0$ scale, as shown in the Eq.~(\ref{dof}).
To reduce the number of partonic degrees of freedom for simplicity and symmetry reasons, we shall make the following assumptions:

\begin{itemize}
    \item $u^{ds}(x) = \bar{u}^{ds}(x)$ and $d^{ds}(x) = \bar{d}^{ds}(x)$, similar to the assumption of $s^{ds}(x) = \bar{s}^{ds}(x) \equiv  s(x)$.
    \item Isospin symmetry for the $u$ and $d$ quarks, so $u^{ds(x)} = d^{ds}(x)$, $\bar{u}^{ds}(x) = \bar{d}^{ds}(x)$.
    \item The DS components of $u$ and $d$ quarks are proportional to the $s$ quarks, i.e.
\end{itemize}

\begin{equation}
    u^{ds}(x) = \bar{u}^{ds}(x) = d^{ds}(x) = \bar{d}^{ds}(x) = Rs(x).
\label{eq:ds_strange}
\end{equation}

\begin{itemize}
    \item We define $u^{cs}(x) = \bar{u}^{cs}(x)$ and $d^{cs}(x) = \bar{d}^{cs}(x)$ so that the valence up- and down-quark PDFs are defined as $u^{v} \equiv u^{v+cs}- \bar{u}^{cs}$. and similar for $d$. We note that  
    this is not the same as the usual definition  $q^v = q - \bar{q}$.     
    The latter has certain conceptual difficulties, such as the strange quark will be a part of the valence when $s^{ds} \neq \bar{s}^{ds}$. It is shown that  next-to-next-to leading order (NNLO)  calculation can perturbatively generate $s^{ds} \neq \bar{s}^{ds}$, and they can also differ in the wavefunction at the $Q_0$ scale, about 1 GeV. Also, the conventional definition has an ambiguity in entangling the $u^v$ and $d^v$ in evolution~\cite{Liu:2017lpe} when $q^{ds} \neq \bar{q}^{ds}$. We further assume that $u^{cs}$ and $d^{cs}$ have the same small-$x$ and large-$x$ behaviors, as the fraction of parton momentum ($x$) inside the proton goes to 0 or 1 limit, respectively. 
     
\end{itemize}

We denote the PDFs global fit with the CS and DS separation specified above as the CT18CSpre fit.
In the above-mentioned third assumption, we have made an ansatz that the $u^{ds}$ and $d^{ds}$ distributions are proportional to the $s^{ds}$ distribution. 
The value of the ratio $R$ is taken from that calculated on the lattice QCD, 

\begin{equation}
    \frac{1}{R} = <x>_{s+\bar{s}}/<x>_{\bar{u}+\bar{d}}(\text{DI}) = 0.795(79)(77) \text{ at 2 GeV~\cite{Liang:2019xdx}, }
\end{equation}
where $<x>_{\bar{u}+\bar{d}}(\text{DI})$ is the momentum fraction carried by $u^{ds}+\bar{u}^{ds}$ (or $d^{ds}+\bar{d}^{ds}$) in the disconnected insertion.

To facilitate this lattice QCD prediction in the CT18CSpre fit, we need to convert the $R$ value evaluated at 2 GeV to 1.3 GeV, the initial scale from which the CT18CSpre PDFs are evolved according to the DGLAP evolution equations. This is done by applying the matching coefficients presented in Ref.~\cite{Yang:2018nqn}, which yields $1/R = 0.822(69)(78)$ at 1.3 GeV.

The distinguishing feature of CS and DS lies in their characteristic small-$x$ behavior.
Since the DS component
can have Pomeron exchanges, we assume in this study that  $q^{ds}(x), \bar{q}^{ds} {}_{\stackrel{\longrightarrow}{x \rightarrow 0}} \, x^{-1}$ for $q = u,d,s$.
In Regge theory, the small-$x$ behavior of $q^{cs}$ and $\bar{q}^{cs}$,
being in the flavor non-singlet connected insertions, are dominated by the reggeon exchanges. To explore its most probable small-$x$ behavior, we have performed a Lagrangian multiplier scan with the ansatz noted above, while allowing all the other fitting parameters to float~\cite{Hou:2019efy}, and found that 
the choice of $q^{cs}(x) = \bar{q}^{cs} {}_{\stackrel{\longrightarrow}{x \rightarrow 0}} \, x^{0}$, for $q = u,d$, can lead to a reasonable fit to the global data set used by the CT18 fits.

\section{Results} \label{sec:results}

\begin{figure}
    \centering
    \includegraphics[width=0.45\textwidth]{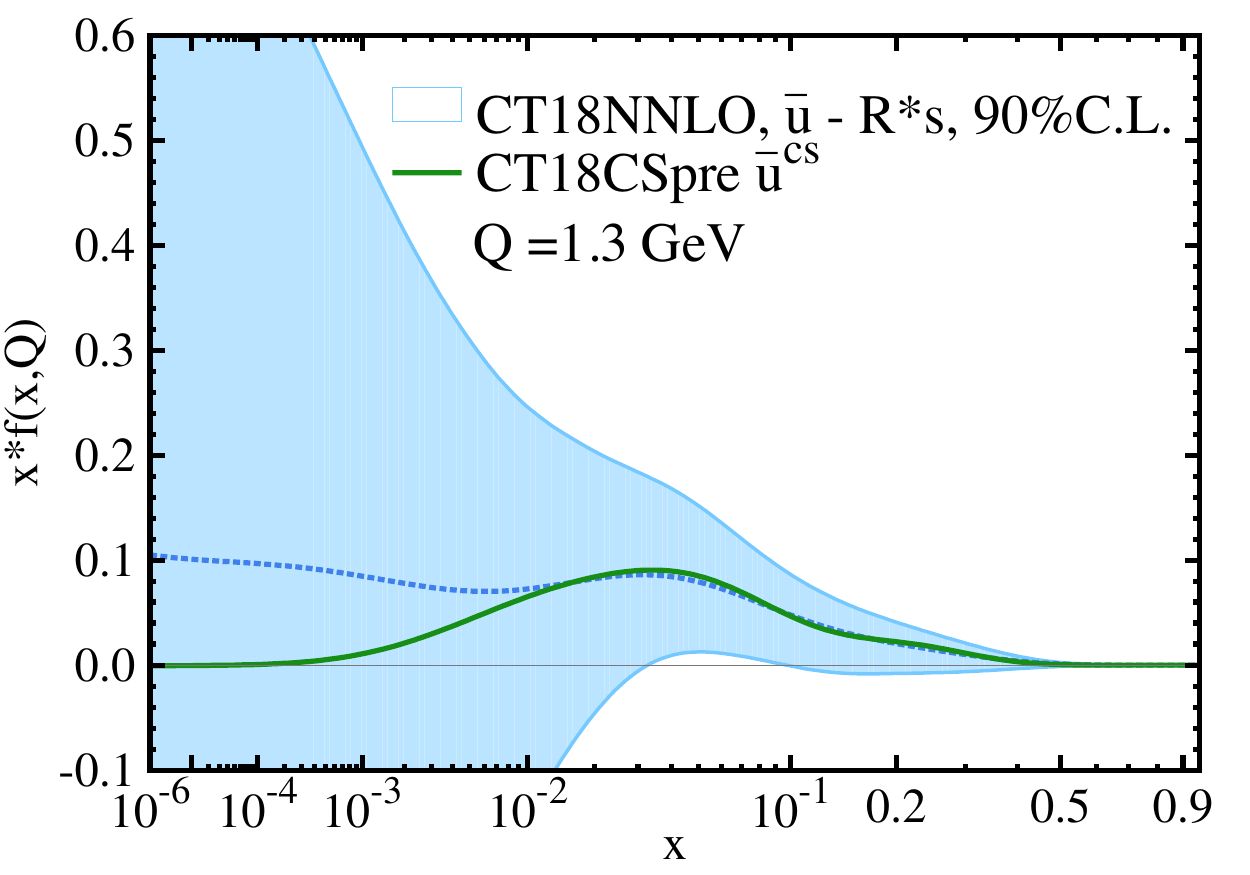} 
    \includegraphics[width=0.45\textwidth]{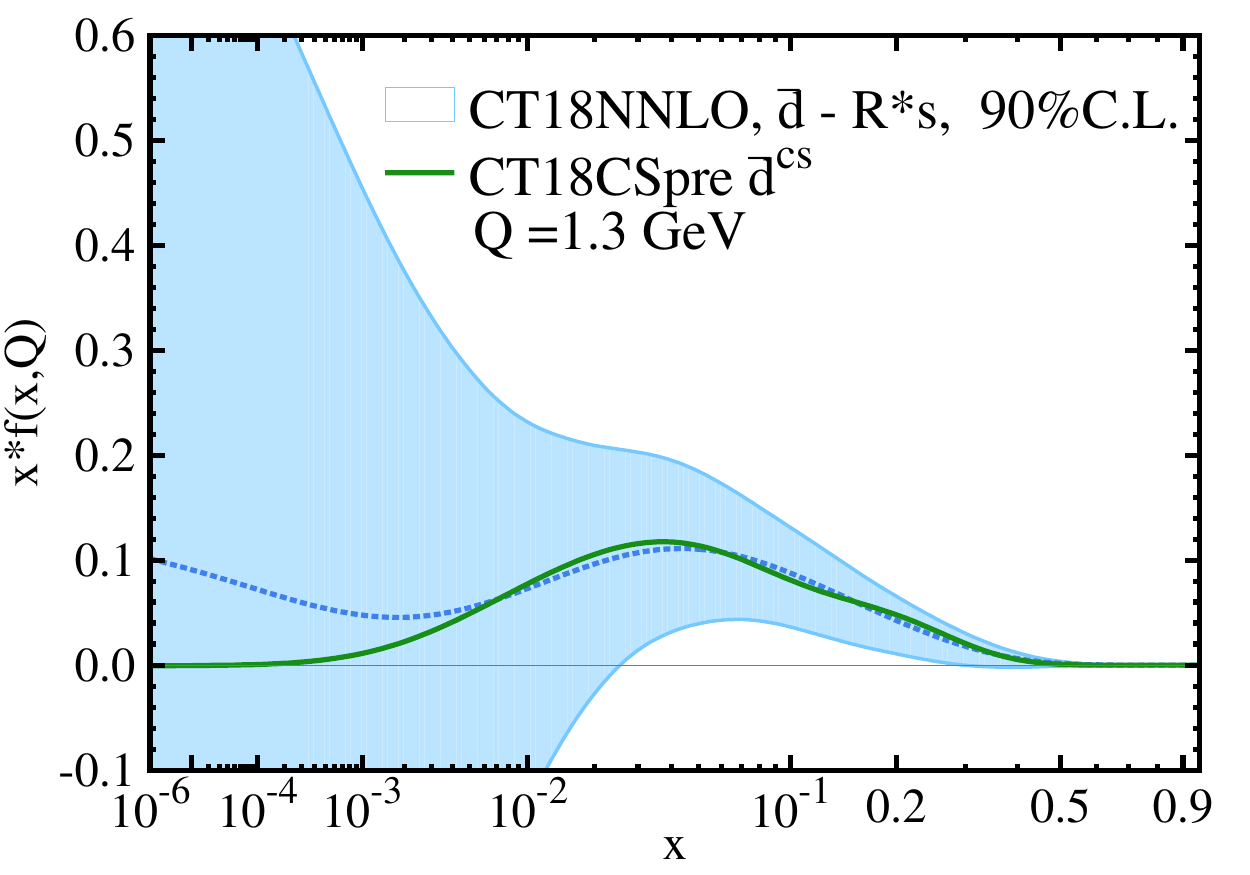} 
    \includegraphics[width=0.45\textwidth]{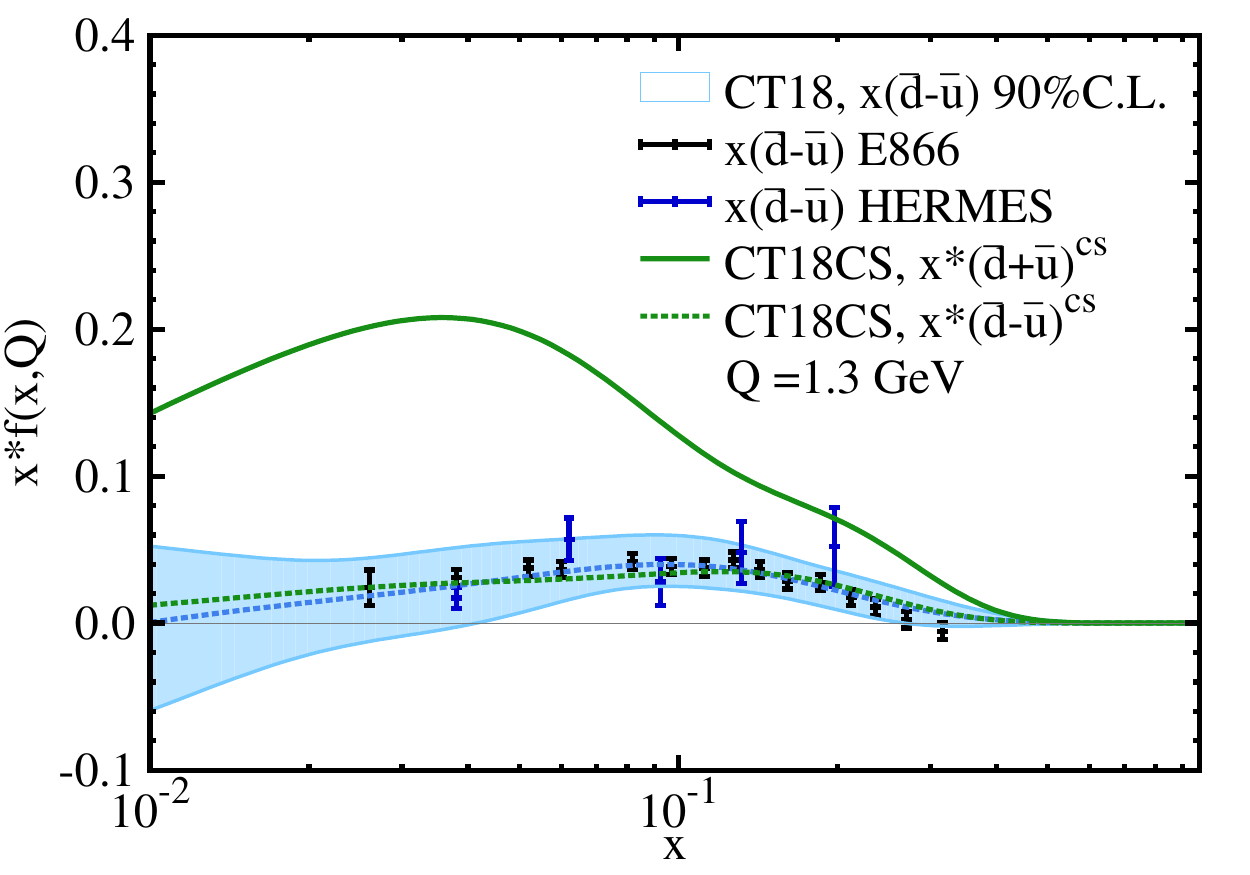} 
    \caption{
    	The upper panel shows the CS components of the up- and down- antiquark distributions in CT18CSpre (green curves), which are compared to the prediction of CT18NNLO (blue dashed lines for central values and blue regions for the error bands at $90\%$ confidence level) on $(\bar u - R*s)$ and $(\bar d - R*s)$, respectively. 
    	The lower panel compares $x(\bar{d}-\bar{u})$ of CT18NNLO and 
    	$x(\bar{d}^{cs}-\bar{u}^{cs})$ of CT18CSpre to the  NuSea E866~\cite{NuSea:2001idv} and HERMES data~\cite{HERMES:2006jyl}, which were obtained from a leading order analysis. We also plot $x(\bar{u}^{cs} + \bar{d}^{cs})$ in CT18CSpre, for comparison.
    \label{fig:PDFs_v+cs}}
\end{figure} 

\begin{figure}
	\centering
	\includegraphics[width=0.45\textwidth]{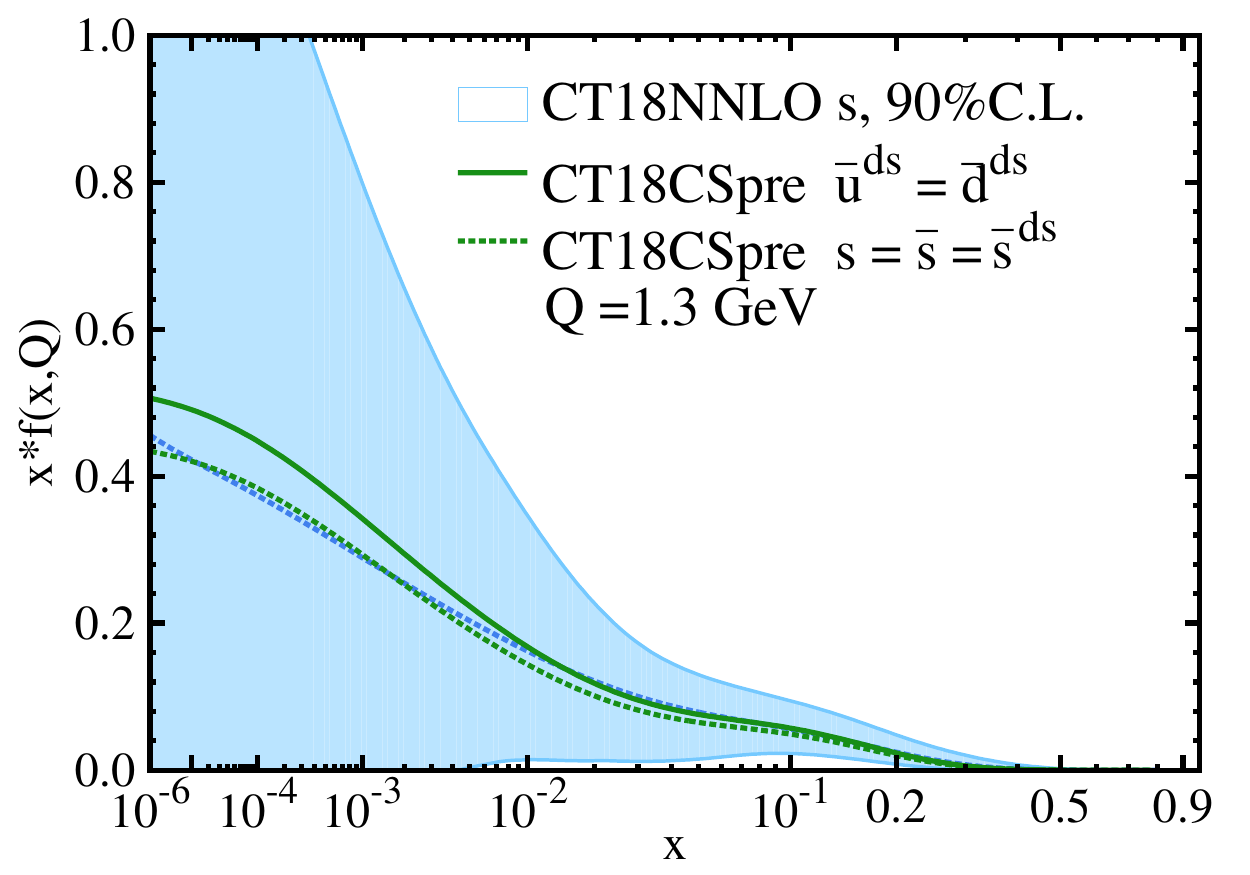} 
	\includegraphics[width=0.45\textwidth]{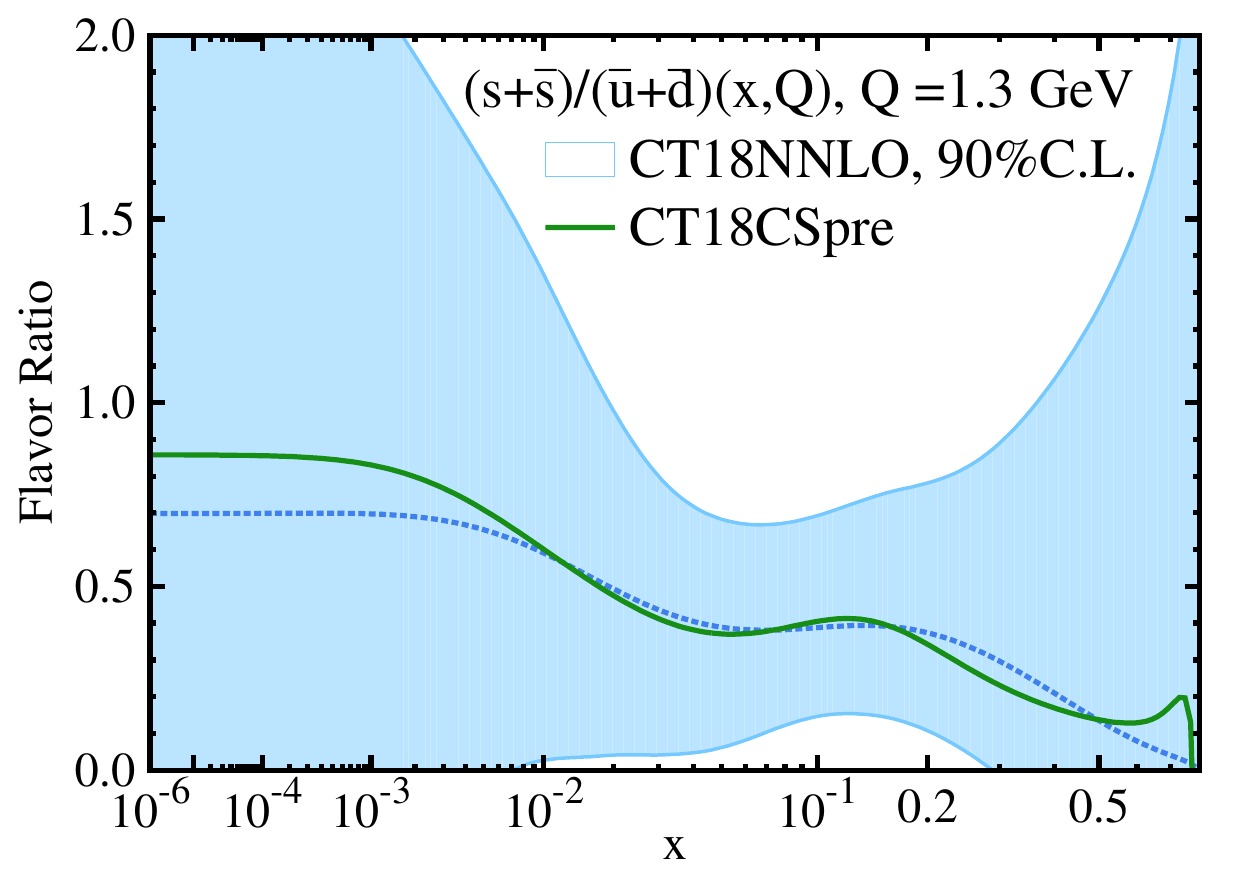} 
	\caption{ Similar to Fig. \ref{fig:PDFs_v+cs}, but for the DS components of $\bar{u}$, $\bar{d}$ and $s$ distributions. The ratio plot of $(s+{\bar s})/({\bar u} +{\bar d})$ is also shown on the right panel at $Q=1.3$ GeV. For CT18CSpre, $\bar{u}$ and $\bar{d}$ include both the CS and DS components, while $s$ only has DS component. 
		\label{fig:PDFs_ds}}
\end{figure} 
We found that the qualities of fits for CT18CSpre and the standard CT18NNLO are comparable. The CT18CSpre, as an alternative parametrisation of CT18, has a total $\chi^2_{\text{CT18CSpre}} = 4299$, which is slightly higher than the standard CT18NNLO value $\chi^2_{\text{CT18}} = 4292$. Considering the CT18 data sets containing 3681 data points totally, both CT18CSpre and CT18NNLO are fitted well. Below, we compare a few interesting features of the fits.
 
Plotted in the upper panel in Fig. 2 are the fitted $x \bar{u}^{cs}$ and $x\bar{d}^{ds}$ in CT18CSpre at Q = 1.3 GeV, which are compared with $x(\bar{u} - R \times \bar{s})(x)$  and $x(\bar{u} - R \times \bar{s})(x)$  in CT18NNLO, respectively.
The lower panel gives $x (\bar{d}(x) - \bar{u}(x))$ in comparision with experiments. 
We also show $x(\bar{d}^{cs}(x) + \bar{u}^{cs}(x))$ which has not been obtained in previous global fits.
The CT18CSpre and CT18NNLO predictions of $x (\bar{d}(x) - \bar{u}(x))$ are in agreement with the NuSea E866 data \cite{NuSea:2001idv} and the HERMES data \cite{HERMES:2006jyl}. The non-vanishing feature of the $x (\bar{d}(x) - \bar{u}(x))$ distributions reflects the violation of the Gottfried sum rule, which motivated this study. Particularly, we attribute the difference in the up- and down-antiquark distributions solely to the difference in their CS components, since their DS components are the same, i.e. $\bar{u}^{ds} = \bar{d}^{ds}$. 
For completeness, we also compare the  disconnected sea distributions in Fig. \ref{fig:PDFs_ds}. The shapes of $\bar{u}^{ds}$, $\bar{d}^{ds}$ and $s^{ds}=s$ are strongly correlated, as implied by the ansatz in Eq. (\ref{eq:ds_strange}).
For the central values of strange PDFs,  CT18CSpre and CT18NNLO show good agreement, particularly in the small-$x$ region. 
The ratio plot of $(s+{\bar s})/({\bar u} +{\bar d})$, evaluated at $Q=1.3$ GeV, shows that CT18CSpre prefers a somewhat larger value than CT18 in the small-$x$ region, where the PDF uncertainty remains to be quite large. 
When $x \ge 10^{-2}$, the ratio starts to dip. This is because $\bar{u}^{cs} + \bar{d}^{cs}$ begins to show up. Here $\bar{u} = \bar{u}^{cs} + \bar{u}^{ds}$ and the same for $\bar{d}$.

Knowledge of the integrated PDF Mellin moments has long been of interest, both for their phenomenological utility,
and for their relevance to lattice QCD computations of hadronic structure.
Separating the CS and DS components of sea quark PDFs allows direct comparison between lattice calculations and global analysis for each parton degree of
freedom.

In Table \ref{tb:moment_1}, the second moments $\langle x \rangle$ for the CT18CSpre at the initial 1.3 GeV scale are compared to their CT18NNLO values, when possible. 
For those partonic degrees of freedom that can be compared, the CT18CSpre and CT18NNLO provide close predictions. 

In the future,  global analyses should incorporate the extended evolution equations~\cite{Liu:2017lpe} where the connected sea and the disconnected sea are evolved separately so that they will remain
separated at all energy scale for better and more detailed delineation of the PDF degrees of freedom.

\begin{table}[htpb]
\begin{center}
\begin{tabular}{l|cccccccc}
 & $u^{v}$ & $d^{v}$ & $\bar{u}^{cs}$ & $\bar{d}^{cs}$ & $\bar{u}^{sea}$ & $\bar{d}^{sea}$ & $s+{\bar s}$ & gluon \\ \hline
 CT18CSpre & 0.323  & 0.136  & 0.013 & 0.021 & 0.029 & 0.037 & 0.032 & 0.386 \\
 CT18NNLO & 0.325 & 0.134 & - & - & 0.028 & 0.036 & 0.027 & 0.385
\end{tabular}
\end{center}
\caption{\label{tb:moment_1} The second moments $\langle x \rangle$ for CT18CSpre and CT18NNLO at $1.3$ GeV. Here, we define  
$\bar{q}^{sea} \equiv \bar{q}^{cs}+\bar{q}^{ds}$, with $q=u$ or $d$, and $s \equiv {s}^{ds}$ for CT18CSpre.
}
\end{table}

 \section{The NuSea and SeaQuest data}
  \label{sec:production}

\begin{figure} [h]
	\centering
	\includegraphics[width=0.45\textwidth]{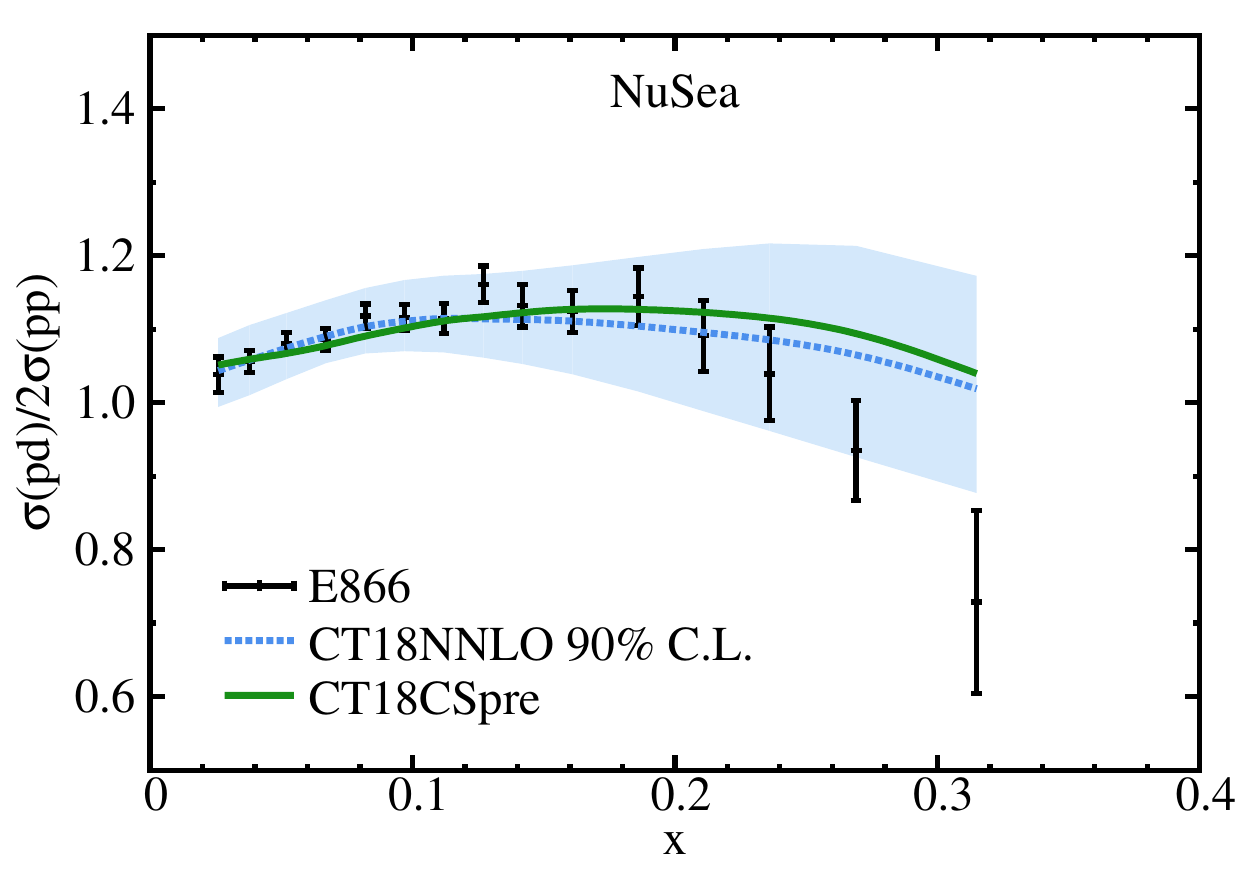} 
	\includegraphics[width=0.45\textwidth]{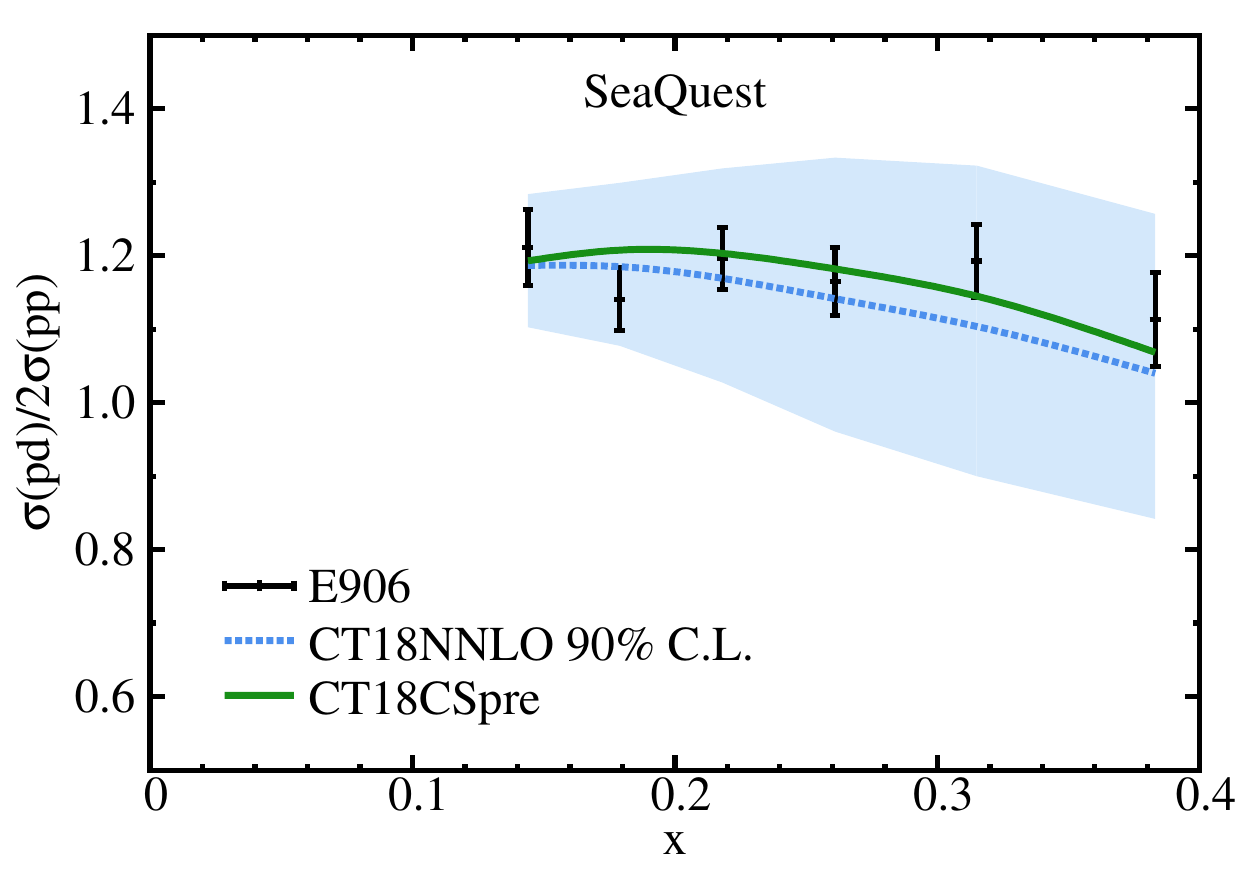} 
	\caption{Comparisons of predictions by CT18CSpre (green curves) and CT18NNLO (blue curves) to the NuSea E866~\cite{NuSea:2001idv} and  SeaQuest E906~\cite{SeaQuest:2021zxb} data, respectively.
		\label{fig:E866_E906}}
\end{figure} 

The NuSea E866 \cite{NuSea:2001idv} and the SeaQuest E906 \cite{SeaQuest:2021zxb} experiments are particularly sensitive to the ratio of $\bar u$ and $\bar d$ PDFs in the intermediate to large-$x$ region, where the CS components dominate. 
As shown in Fig. \ref{fig:E866_E906}, the predictions by the CT18CSpre and CT18NNLO PDFs are mostly consistent with NuSea and SeaQuest experimental data, except for the last bin of the NuSea data with $x > 0.3$. Comparing to the CT18NNLO predictions, the CT18CSpre predictions are slightly shifted upwardly, in better agreement with the SeaQuest data. We note that the SeaQuest data were not included in either the CT18NNLO or CT18CSpre fits.

\section{Conclusions} \label{summary}

The Euclidean path-integral formulation of the hadronic tensor of the nucleon uncovered that there are two kinds of sea partons, one is the
connected sea and the other disconnected sea. 
To isolate the disconnected component of sea quarks, we assume that $u^{ds}$ and $d^{ds}$ are proportional to that the strange, i.e. $u^{ds}(x) = d^{ds} = Rs(x)$ (see Eq.~(\ref{eq:ds_strange})), where R is taken from the lattice calculation of the ratio of the strange momentum fraction to that of the ${\bar u}$ or ${\bar d}$ quark in the disconnected insertion.
A new global fit with the separate degrees of freedom of connected and disconnected sea partons results in CT18CSpre, which is consistent with the nominal CT18NNLO in the measurement of total $\chi^2$.
As expected, the CS components are dominating the total sea quark distributions in the intermediate to large $x$ region, while the DS pieces contribute predominantly in the small-$x$ region. 
In general, the parton degrees of freedom in CT18CSpre are more than those in CT18NNLO.
We have compared various flavor PDFs of CT18CSpre and CT18 to the extent that they can be compared. For example, we can combine ${\bar u}^{cs}$ and ${\bar u}^{ds}$ in CT18CSpre to $\bar u$ in CT18NNLO, etc. 

Separating the CS and DS components of sea quark PDFs allows direct comparison between lattice calculations and global analysis for each parton degree of
freedom.
In the future,  global analyses should incorporate the extended evolution equations~\cite{Liu:2017lpe} where the connected sea and the disconnected sea are evolved separately so that they will remain
separated at all energy scale for better and more detailed fits of the PDF degrees of freedom.

\section{Acknowledgment}
The authors are indebted to  J.C. Peng, J.W. Qiu, and Y.B. Yang for insightful discussions.  This work is partially support by the U.S. DOE grant DE-SC0013065 and DOE Grant No.\ DE-AC05-06OR23177 which is within the framework of the TMD Topical Collaboration.
This research used resources of the Oak Ridge Leadership Computing Facility at the Oak Ridge National Laboratory, which is supported by the Office of Science of the U.S. Department of Energy under Contract No.\ DE-AC05-00OR22725. This work used Stampede time under the Extreme Science and Engineering Discovery Environment (XSEDE), which is supported by National Science Foundation Grant No. ACI-1053575.
We also thank the National Energy Research Scientific Computing Center (NERSC) for providing HPC resources that have contributed to the research results reported within this paper.
We acknowledge the facilities of the USQCD Collaboration used for this research in part, which are funded by the Office of Science of the U.S. Department of Energy.
The work at MSU is partially supported by the U.S.~National Science Foundation
under Grant No.~PHY-2013791.
C.-P.~Yuan is also grateful for the support from
the Wu-Ki Tung endowed chair in particle physics.


\nolinenumbers

\end{document}